\documentclass{emulateapj} 
 
\usepackage{graphicx}

\newcommand{\ts}{\thinspace}
\newcommand{\etal}{\mbox{et~al.}}

\newcommand{\Lsol}{\mbox{$L_{\odot}$}}

\def\deg      {{\ifmmode^\circ\else$^\circ$\fi}} 

 \shorttitle{S-COSMOS}
 \shortauthors{Sanders et al.}
 \slugcomment{ApJS:  accepted 01/09/07}
 
\begin{document}
 
 
 \title{S-COSMOS: The {\it Spitzer} Legacy Survey of the HST-ACS 2{\ts}\sq\deg\  COSMOS Field$^*$I: survey strategy and first analysis.}

 \author{ 
 D. B. Sanders\altaffilmark{1},
 M. Salvato\altaffilmark{2},
H. Aussel\altaffilmark{1,3},
O. Ilbert\altaffilmark{1},
N. Scoville\altaffilmark{2,4},
J. A. Surace\altaffilmark{5},
D. T. Frayer\altaffilmark{5},
K. Sheth\altaffilmark{5},
G. Helou\altaffilmark{2,5},
T. Brooke\altaffilmark{2}, 
B. Bhattacharya\altaffilmark{5},
L. Yan\altaffilmark{5},
J. S. Kartaltepe\altaffilmark{1},
J. E. Barnes\altaffilmark{1},
A. W. Blain\altaffilmark{2},
D. Calzetti\altaffilmark{6},
P. Capak\altaffilmark{2},
C. Carilli\altaffilmark{7},
C. M. Carollo\altaffilmark{8},
A. Comastri\altaffilmark{9},
E. Daddi\altaffilmark{10},
R. S. Ellis\altaffilmark{2},
M. Elvis\altaffilmark{11},
S. M. Fall\altaffilmark{6},
A. Franceschini\altaffilmark{12},
M. Giavalisco\altaffilmark{6},
G. Hasinger\altaffilmark{13},
C. Impey\altaffilmark{14},
A. Koekemoer\altaffilmark{6},
O. Le F\`evre\altaffilmark{15},
S. Lilly\altaffilmark{8},
M. C. Liu\altaffilmark{1,16},
H. J. McCracken\altaffilmark{17,18},
B. Mobasher\altaffilmark{6},
A. Renzini\altaffilmark{12,19},
M. Rich\altaffilmark{20},
E. Schinnerer\altaffilmark{21},
P. L. Shopbell\altaffilmark{2},
Y. Taniguchi\altaffilmark{22},
D. J. Thompson\altaffilmark{23,24},
C. M. Urry\altaffilmark{25},
J. P. Williams\altaffilmark{1}}



\altaffiltext{$\star$}{Based on observations with the NASA/ESA {\em
Hubble Space Telescope}, obtained at the Space Telescope Science
Institute, which is operated by AURA Inc, under NASA contract NAS
5-26555; also based on data collected at : the Subaru Telescope, which is operated by
the National Astronomical Observatory of Japan; the XMM-Newton, an ESA science mission with
instruments and contributions directly funded by ESA Member States and NASA; the 
European Southern Observatory under Large Program 175.A-0839, Chile; Kitt Peak 
National Observatory, Cerro Tololo Inter-American
Observatory, and the National Optical Astronomy Observatory, which are
operated by the Association of Universities for Research in Astronomy, Inc.
(AURA) under cooperative agreement with the National Science Foundation; 
the National Radio Astronomy Observatory which is a facility of the National Science 
Foundation operated under cooperative agreement by Associated Universities, Inc ; 
and and the Canada-France-Hawaii Telescope with MegaPrime/MegaCam operated as a
joint project by the CFHT Corporation, CEA/DAPNIA, the National Research
Council of Canada, the Canadian Astronomy Data Centre, the Centre National
de la Recherche Scientifique de France, TERAPIX and the University of
Hawaii.}  
\altaffiltext{1}{Institute for Astronomy, 2680 Woodlawn Dr., University of Hawaii, Honolulu, Hawaii, 96822; sanders@ifa.hawaii.edu}
\altaffiltext{2}{California Institute of Technology, MS 105-24, 1200 East California Boulevard, Pasadena, CA 91125}
\altaffiltext{3}{CNRS, AIM -- Unit\'e Mixte de Recherche CEA -- CNRS -- Universit\'e Paris VII -- UMR n$^\circ$ 7158, 91191 Gif-sur-Yvette, France}
\altaffiltext{4}{Visiting Astronomer, Univ. Hawaii, 2680 Woodlawn Dr., Honolulu, HI, 96822}
\altaffiltext{5}{Spitzer Science Center, MS 314-6, California Institute of Technology, Pasadena, CA 91125}
\altaffiltext{6}{Space Telescope Science Institute, 3700 San Martin Drive, Baltimore, MD 21218}
\altaffiltext{7}{National Radio Astronomy Observatory, P.O. Box 0, Socorro, NM  87801-0387}
\altaffiltext{8}{Department of Physics, ETH Zurich, CH-8093 Zurich, Switzerland}
\altaffiltext{9}{INAF-Osservatorio Astronomico di Bologna, via Ranzani 1, 40127 Bologna, Italy}
\altaffiltext{10}{National Optical Astronomy Observatory, P.O. Box 26732, Tucson, AZ 85726}
\altaffiltext{11}{Harvard-Smithsonian Center for Astrophysics, 60 Garden Street, Cambridge, MA 02138}
\altaffiltext{12}{Dipartimento di Astronomia, Universitˆ di Padova, vicolo dell'Osservatorio 2, I-35122 Padua, Italy}
\altaffiltext{13}{Max Planck Institut f\"ur Extraterrestrische Physik,  D-85478 Garching, Germany}
\altaffiltext{14}{Steward Observatory, University of Arizona, 933 North Cherry Avenue, Tucson, AZ 85721}
\altaffiltext{15}{Laboratoire d'Astrophysique de Marseille, BP 8, Traverse du Siphon, 13376 Marseille Cedex 12, France}
\altaffiltext{16}{Alfred P. Sloan Research Fellow}
\altaffiltext{17}{Institut d'Astrophysique de Paris, UMR7095 CNRS, Universit\`e Pierre et Marie Curie, 98 bis Boulevard Arago, 75014 Paris, France}
\altaffiltext{18}{Observatoire de Paris, LERMA, 61 Avenue de l'Observatoire, 75014 Paris, France}
\altaffiltext{19}{European Southern Observatory, Karl-Schwarzschild-Str. 2, D-85748 Garching, Germany}
\altaffiltext{20}{Department of Physics and Astronomy, University of California, Los Angeles, CA 90095}
\altaffiltext{21}{Max Planck Institut f\"ur Astronomie, K\"onigstuhl 17, Heidelberg, D-69117, Germany}
\altaffiltext{22}{Physics Deptartment, Graduate School of Science, Ehime University, 2-5 Bunkyou, Matuyama, 790-8577, Japan}
\altaffiltext{23}{Caltech Optical Observatories, MS 320-47, California Institute of Technology, Pasadena, CA 91125}
\altaffiltext{24}{Large Binocular Telescope Observatory, University of Arizona, 933 N. Cherry Ave.,
   Tucson, AZ  85721}
\altaffiltext{25}{Department of Astronomy, Yale University, P.O. Box 208101, New Haven, CT 06520-8101}

 \begin{abstract}
The {\it Spitzer}-COSMOS survey (S-COSMOS) is a Legacy program (Cycles 2+3) 
designed to carry out a uniform deep survey of the full 2{\ts}\sq\deg\  COSMOS field in 
all seven {\it Spitzer} bands (3.6, 4.5, 5.6, 8.0, 24.0, 70.0, 160.0{\ts}$\mu$m). 
This paper describes the survey parameters, mapping strategy, data reduction procedures, 
achieved sensitivities to date, and the complete data set for future reference.
We show that the observed infrared backgrounds in the S-COSMOS field are   
within 10\% of the predicted background levels.  The fluctuations in the background at 24{\ts}$\mu$m have 
been measured and do not show any significant contribution from cirrus, as expected. 
In addition, we report on the number of asteroid detections in the low galactic latitude COSMOS field. 
We use the Cycle 2 S-COSMOS data to determine preliminary number counts, and compare our  
results with those from previous  {\it Spitzer} Legacy surveys (e.g. SWIRE, GOODS).  
The results from this ``first analysis" confirm that the S-COSMOS survey will have 
sufficient sensitivity with IRAC to detect $\sim L^*$ disks and spheroids out to $z \gtrsim 3$, 
and with MIPS to detect ultraluminous starbursts and AGN out to $z \sim 3$ at 24{\ts}$\mu$m and out to 
$z \sim 1.5-2$ at 70{\ts}$\mu$m and 160{\ts}$\mu$m.  
 \end{abstract}

 
 

 
 \section{Introduction}
 
  The Cosmic Evolution Survey (COSMOS), covering 2 \sq\deg, 
 is the first HST survey specifically designed to thoroughly probe the evolution of 
 galaxies, AGN and dark matter in the context of their cosmic 
 environment (large scale structure -- LSS) in a contiguous field that samples a 
 volume in the high redshift universe approaching that sampled locally by the 
 Sloan Digital Sky Survey (SDSS).  An overview of the COSMOS 
 project, including a full characterization of the field and a description of the 
 vast amount of multi-wavelength data  which has been, and will be, assembled 
 is given by \citet{sco06}.  
 
 The {\it Spitzer}-COSMOS (S-COSMOS) deep IRAC and MIPS data are
 critical to address two major goals of the COSMOS survey: the
 stellar-mass assembly of galaxies (primarily from IRAC) and a full
 accounting of the luminosity from dust-embedded sources such as
 merging starburst galaxies and AGN (MIPS and IRAC).  IRAC measures
 the light from long-lived stars in galaxies over the redshift range
 $0.5 < z < 6$ \citep[e.g.,][]{mob05}. With morphological information
 from the HST-ACS imaging, we will be able to determine how the
 mass-assembly depends on morphological type, environment and
 redshift, in addition to other properties such as X-ray and radio
 emission.  IRAC data are also critical for determining accurate
 photometric redshifts for galaxies at $z > 1$ \citep[e.g.,][]{row05},
 where the peak of the stellar light has been redshifted into the IRAC
 bands.  MIPS observations allow a determination of the total star
 formation rate (SFR) without requiring large corrections for
 extinction in galaxies, and will be used to quantify the connection
 between the SFR and the merger/interaction rate
 \citep[e.g.,][]{kar06,kam06} and the dependence on clustering
 environment (derived from spectroscopic and photometric
 redshifts). Deep COSMOS radio images  \citep{sch06} will also be used
 to provide a secondary calibration of SFRs.  In addition, both IRAC
 and MIPS probe the wavelength range where obscured AGN have much of
 their bolometric luminosity.  If the AGN are heavily-obscured they
 may be missing from the XMM-COSMOS survey, which is the primary basis
 for AGN selection \citep{imp06}, but they will appear in the catalogs
 of infrared point sources.  Mid-infrared selection has been shown to
 be $\sim${\ts}60\% more efficient in selecting highly obscured AGN
 \citep{pol06}.  The COSMOS field, with its size, deep mid-IR coverage
 and extensive UV/X-Ray/Radio/Optical data set offers the possibility
 to both significantly improve the constraint on the surface density
 of obscured AGN and to improve our understanding of their spectral
 energy distributions (SEDs).
 
 This paper presents the infrared properties of the COSMOS field, summarizes  our mapping 
 strategy, and presents the initial quick-look results from our Cycle 2 Legacy observations, all 
 of which were completed during the 2005 December - 2006 January, visibility window.  
 Section \ref{fieldcharacterization} briefly summarizes the general properties of the COSMOS Field
 (e.g. location, expected infrared backgrounds),  and  \S \ref{observingstrategy} lays out the mapping 
 strategy used to obtain coverage of the full field with both the IRAC and MIPS cameras.  The methods 
 used to reduce our Cycle 2 data are presented in \S \ref{datareduction}.  Section \ref{prelimanalysis} 
 presents results from our preliminary analysis of the Cycle 2 data, including measured background levels 
 and sensitivities, preliminary number counts, and asteroid detections.   Our future science goals are summarized in 
 \S \ref{future}.  
 
\section  {Field Characterization}
\label{fieldcharacterization}

\subsection{Field location}

The COSMOS field is equatorial to ensure coverage by all astronomical facilities 
(centered at J2000 RA =10:00:28.6, Dec = +02:12:21.0). Our field was chosen to be devoid 
of bright X-ray, UV, and radio sources. The time requirements for imaging over a total area 
of 2{\ts}\sq\deg\  and ground-based spectra of 50,000 galaxies at $I_{\rm AB} < 25$ mag makes it strategically 
imperative that the field be readily observable now and in the future by all large optical/IR 
telescopes and especially all unique astronomical instruments (e.g.  ALMA, EVLA, SKA, TMT, JCMT/SCUBA2).
 
 \subsection{IR Backgrounds}
 
\begin{figure}[ht]
\epsscale{1.0} 
\plottwo{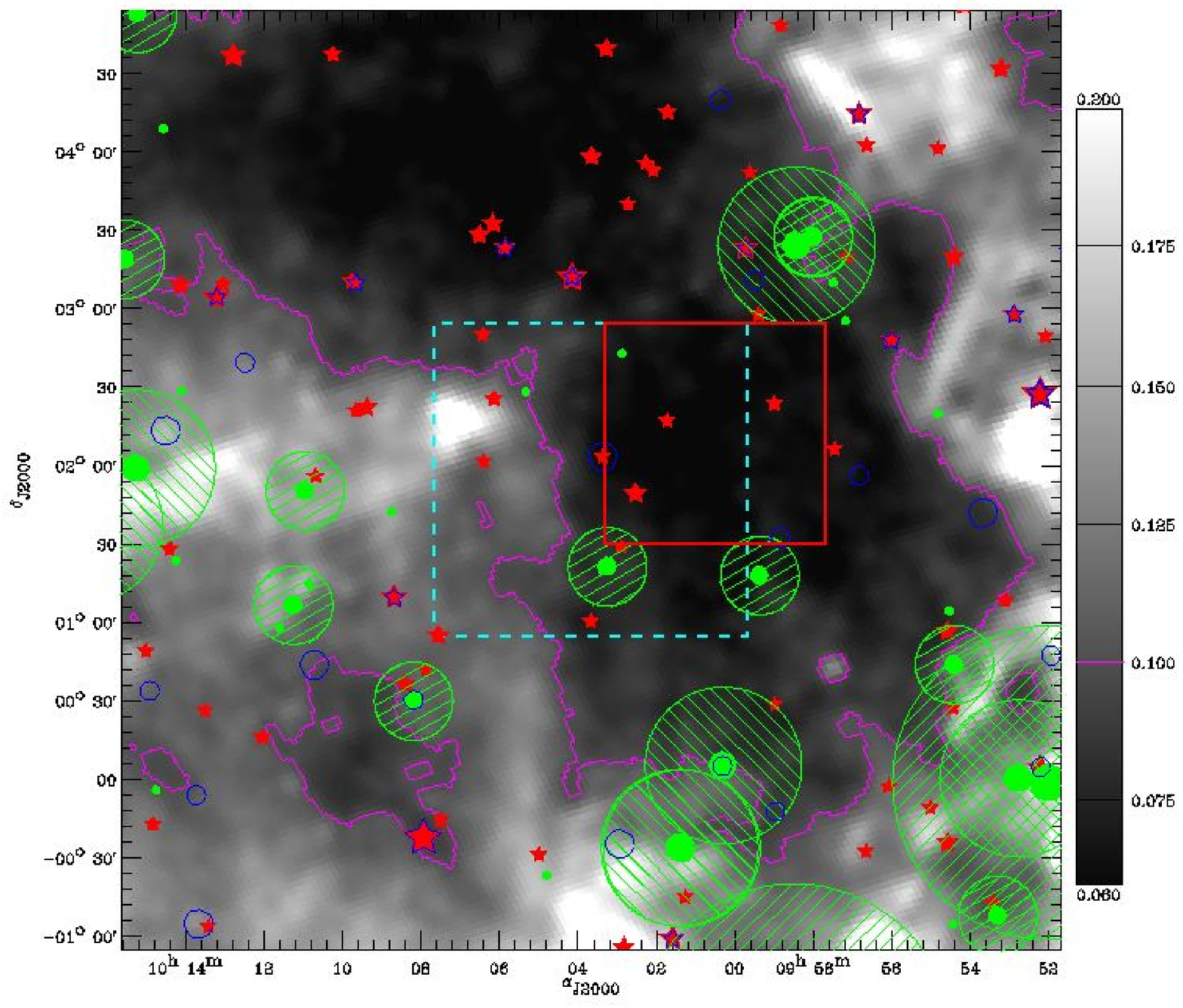}{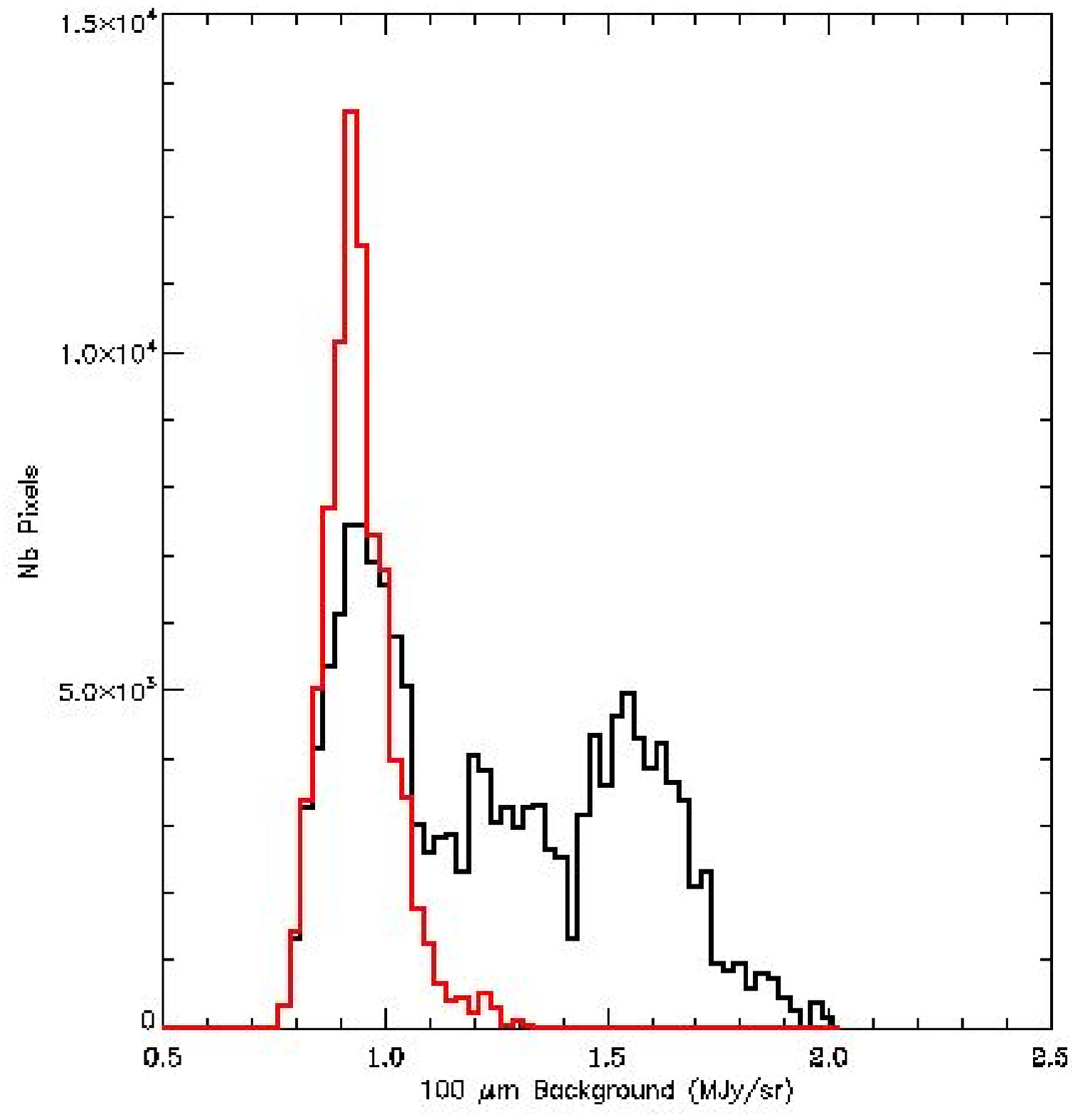}
\caption{{\bf (Left panel:)}\ The layout of the COSMOS Field  on a map of extinction computed from the reddening map of \citet{sch98}.  
The COSMOS Field is indicated by the red rectangle ($1.41^\circ \times 1.41^\circ$) fully enclosing all of the ACS imaging, which has lower left and 
upper right corners (RA,DEC J2000) at  (150.7988\deg,1.5676\deg) and  (149.4305\deg, 2.8937\deg).
The dashed blue box represents the $2^\circ \times 2^\circ$ VVDS field.  Tycho-2 bright stars are indicated by red stars.
Bright NVSS radio sources are indicated by green dots.  Radio sources with $S_{\rm 1.4GHz} > 1$Jy, 0.5-1.0Jy, and 0.25-0.5Jy,  are surrounded 
by an hatched area circle of radius 1.5$^\deg$, 1$^\deg$ and 0.5$^\circ$, respectively.   ROSAT All-Sky Survey Bright 
Source Catalog sources are indicated by blue circles.  The purple contour represents an extinction value of $A_{\rm V} = 0.1$.
\ {\bf (Right panel:)}\  Histogram of the distribution of 100{\ts}$\mu$m {\it IRAS} emission over the entire 2{\ts}\sq\deg\  COSMOS field (red) compared to the larger VVDS field (black).} 
\label{cosmos_field}
\end{figure}

The COSMOS field was chosen to have among the lowest (and most uniform) mean IRAS-100{\ts}$\mu$m 
background level for an equatorial field of its size.   Figure \ref{cosmos_field}  shows the field outline 
(red square in the left panel; see caption for details) superimposed on 
a map of extinction computed from the reddening map of \citet{sch98}, illustrating the apparent 
lack of contamination from foreground Galactic Plane dust clouds, as well as the lack of any strong 
radio and X-Ray sources which could compromise deep radio and X-ray observations of the field.  Figure 
\ref{cosmos_field} also  shows in histogram form the distribution of 100{\ts}$\mu$m {\it IRAS} pixels within the COSMOS field versus 
the larger surrounding VVDS field.  The sharp peak near 0.9 MJy/sr, and the lack of the higher 
background tail that plagues the larger VVDS region (dashed blue box in Figure \ref{cosmos_field}, left panel), 
illustrates the choices made in shifting the final COSMOS field 
center to truly minimize the far-infrared background levels. 

Although lower background fields at higher declination and higher ecliptic latitude will have slightly better sensitivities for an 
equal amount of integration time, these fields suffer the major deficiency that they are not readily 
accessible by all unique facilities Ð most importantly the (E)VLA and future ALMA which are vital 
for multi-wavelength studies and the future 30m optical telescope which is likely to be unique.  
We believe that the accessibility to all instruments far outweighs the relatively small loss ($\sim 35$\%) 
in sensitivity for Spitzer observations of the COSMOS field.

 \section{Observing Strategy}
 \label{observingstrategy}

Because the HST-COSMOS field is close to the ecliptic, the orientation
of the detectors is always the same to within a few degrees.  This
allowed specification of a maximally efficient tiling strategy, since
we could accurately predict the orientation of the detectors.  Neither
of the mapping options available in the Astronomical Observation Template (AOT) allowed us the flexibility
to tile given the fact that the COSMOS ACS data is aligned along the
RA-DEC coordinate axes, so instead the fields have been
placed using the ``cluster offset" option.  

\subsection{Cycle{\ts}2: IRAC-deep Coverage}
\begin{figure}[ht]
\epsscale{0.8} 
\plotone{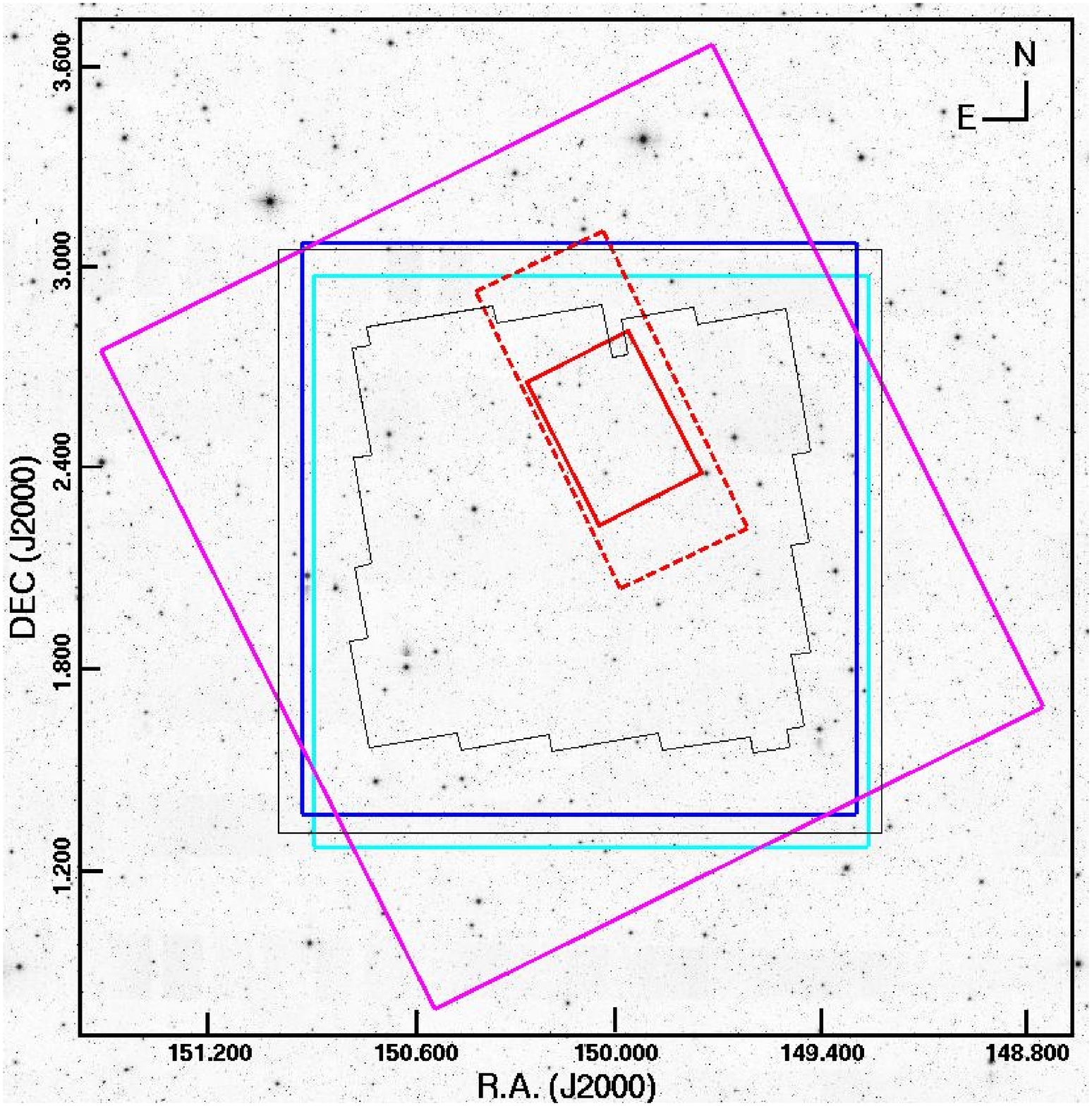}
\caption{Visualization of the S-COSMOS Cycle 2 IRAC and MIPS AOR coverage. 
The background image is from the SDSS.  Thin black square=CFHT/I-band, thick black polygon-ACS/i-band,
 Magenta square=MIPS-shallow, Dashed Red rectangle=MIPS-deep ``Test area" coverage at 24{\ts}$\mu$m, 
 Solid Red rectangle=MIPS-deep  ``Test area" with 
 coverage in all three bands, Blue square=IRAC-3.6{\ts}$\mu$m, 5.4{\ts}$\mu$m,  Cyan square= IRAC-4.5$\mu$m, 8.0{\ts}$\mu$m. Coordinates of the corners of each box are listed in Table \ref{tbl-fieldscoo}.
} 
\label{coverage_cycle2}
\end{figure}

The IRAC camera has two field of views (FOVs), 
each one dedicated to two channels -- one for the 3.6{\ts}$\mu$m and 5.6{\ts}$\mu$m detectors, 
and one for the 4.5{\ts}$\mu$m and 8.0{\ts}$\mu$m detectors.  The two FOVs  are separated on the 
sky by  7$^\prime$.  In order to fully cover the ACS-COSMOS field at nominal depth with 
all four IRAC bands, we had to map a slightly larger region of the sky, as shown 
in Figure \ref{coverage_cycle2}.   

The following four constraints dictated our final choice for the map
tiling: (1) The maximum size of an IRAC Astronomical Observation Request (AOR) is 6{\ts}hrs, and in practice
must be less than 5{\ts}hrs.; \ (2) Because the HST-COSMOS field is near
the ecliptic plane, asteroids were expected be a significant
contaminant.  We required multiple epochs separated by $1-2${\ts}hrs
for asteroid rejection, plus each epoch must have enough redundancy to
allow for cosmic ray rejection;\ (3) Each epoch had to have at least
$4-5$ dithers per point on the sky for adequate redundancy when reducing
the data.  A total of four epochs were observed;  (4) IRAC has 2 FOVs
separated by $7^\prime$.  The orientation of the FOVs was
dictated by the orientation of the spacecraft on the day of the
observation.  Both FOVs had to observe the entire field.

The basic parameters of the tiling are a $4 \times 4$ grid of AORs,
where each AOR is composed of a grid of $5\times5$
IRAC FOVs.  Each AOR in the $4\times4$ grid was repeated sequentially three times, 
resulting in three individual epochs separated by a few hours. 
The east and south edges are trimmed to match the HST-COSMOS
field boundary.  The fields overlap by tens of arcsecs, and are highly
dithered.  
Each epoch was further offset half an array width relative
to its predecessor in order to ensure an even higher number of total dither positions 
in the final coaddition of all epochal data in order to minimize array-position dependent calibration effects. 
Figure \ref{coverage_cycle2}
shows the final area coverage resulting from our tiling strategy for the two IRAC FOVs.  
The total time for the IRAC-deep full-field mapping
observations (all carried out in Cycle 2) was 166{\ts}hrs (see Table \ref{tbl-obssummary}).

\subsection{Cycle{\ts}2: MIPS Coverage}
In Cycle{\ts}2, a total of 58.2{\ts}hrs were allocated for MIPS observations
of the COSMOS field.  The entire field was mapped at a shallow depth
to check for possible cirrus structure within the COSMOS field, and
deep observations over a small $30^\prime \times 20^\prime$ region were carried out to
quantify the ability of Spitzer to integrated down within the COSMOS
field (see solid red line in Figure \ref{coverage_cycle2}).  The data were taken in 
MIPS scan mode with $148^{\prime\prime}$
cross-scan offsets between the forward and return scan legs.  The
shallow observations were taken at the medium scan rate with $1.75^\deg$
scan legs.  These shallow observations cover $1.75^\deg \times  1.97^\deg$ to a
depth of 80{\ts}sec, 40{\ts}sec, and 8{\ts}sec for the 24{\ts}$\mu$m, 70{\ts}$\mu$m, and 160{\ts}$\mu$m arrays,
respectively.

The MIPS-deep ``Test area" observations were taken at the slow scan rate
with $0.5^\circ$ scan legs.  These deep observations cover $0.5^\deg \times 0.33^\deg$ 
to a depth of approximately 3200{\ts}sec, 1560{\ts}sec, 320{\ts}sec for the 24{\ts}$\mu$m, 70{\ts}$\mu$m, and
160{\ts}$\mu$m arrays respectively.  The small ``Test area" was mapped 15 times
using dithers between each map to account for the bad blocks within
the 70{\ts}$\mu$m and 160{\ts}$\mu$m arrays and to spread out the overlapping 24{\ts}$\mu$m
coverage uniformly over the field.  Both forward and reverse scan
mapping were done to help characterize any long-term transients as a
function of scan direction.

\subsection{Cycle{\ts}3:  MIPS-deep Coverage of the Full-Field}
\begin{figure}[ht]
\epsscale{0.8} 
\plotone{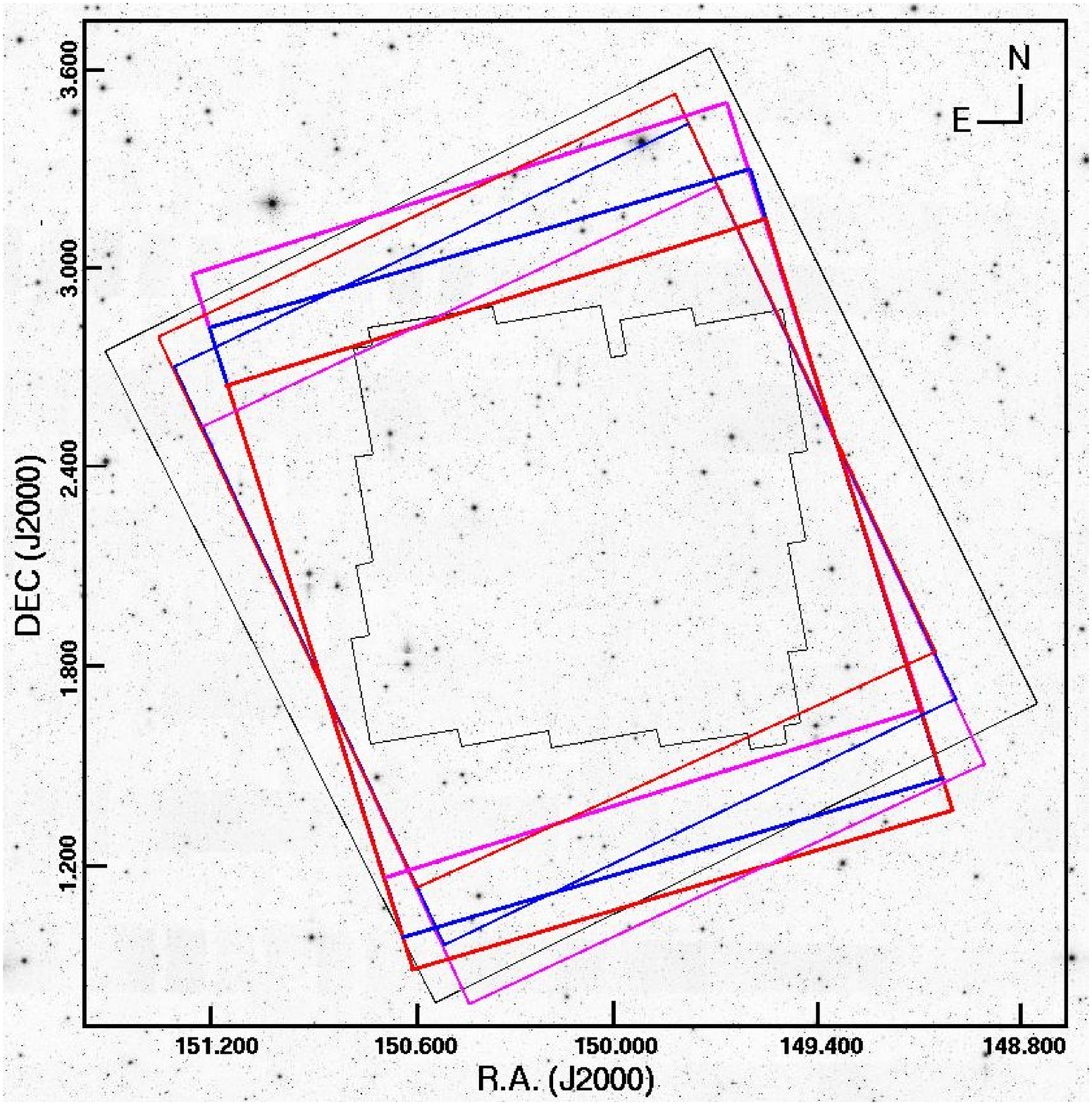}
\caption{Visualization of the Cycle 3 MIPS-deep coverage of the COSMOS field. 
The thin black ACS and MIPS-shallow Cycle 2 observations are as in Figure 2.  The blue, magenta, and red boxes are 
respectively, the 24{\ts}$\mu$m, 70{\ts}$\mu$m, and 160{\ts}$\mu$m observations to be obtained in Cycle 3.  The two ``sets" of boxes 
represent the two different spacecraft orientations for the two epochs. Coordinates of the corners of each box are listed in 
Table \ref{tbl-fieldscoo}. 
} 
\label{coverage_cycle3}
\end{figure}

In Cycle{\ts}3, 396.2{\ts}hrs have been allocated for deep MIPS observations
covering the full COSMOS field.  These observations will be taken in
two epochs (2007, January and 2007, May) to help minimize the
zodical background and thus to maximize the achieved sensitivity.  The observations 
during the two epochs will of necessity have different instrumental orientations 
(see Figure \ref{coverage_cycle3}).  The field
will be fully mapped 13 times at the slow scan rate with $148^{\prime\prime}$
cross-step offsets between the adjacent $1.5^\deg$ scan legs.  The field
will also be mapped once at the medium scan rate to confirm the consistency of the 
calibration between the Cycle{\ts}2 and Cycle{\ts}3 data sets for the same integration time.   
The Cycle{\ts}3 observations will cover $1.5^\deg \times 1.64^\deg$ 
to a depth of approximately 2800{\ts}sec, 1400{\ts}sec, and 280{\ts}sec for the
24{\ts}$\mu$m, 70{\ts}$\mu$m, and 160{\ts}$\mu$m arrays respectively (see Figure  \ref{coverage_cycle3}).

\section{Data Reduction (Cycle{\ts}2)}
\label{datareduction}
\subsection{IRAC}

The S-COSMOS data were initially processed by the Spitzer Science
Center\footnote{http://ssc.spitzer.caltech.edu/} (SSC). The SSC
provides the basic calibrated data (BCD) product, which are raw scientific
exposures that are flux calibrated and corrected for
well-understood instrumental signatures. These include dark
subtraction, linearization, and flat-fielding.
After receipt of the BCD images, S-COSMOS further processes them using
a data pipeline adapted from the SWIRE project and described by \citet{sur07}.
This pipeline fixes numerous additional instrumental effects such as
muxbleed, jailbarring, muxstripe, banding, and column
pull-up and pull-down.  It applies a frame-by-frame background correction for
fluctuating bias levels. It corrects both long and short-term
image latents, and masks straylight and filter ghosts. Finally, it
applies a correction for the known array-position dependent
calibration gain factors. 

One caveat of this processing, which is true for all IRAC observations, is that the
background on 5\arcmin\ and larger scales is fixed to a COBE-derived background
model.  Operated in a shutterless mode, IRAC cannot provide an absolute measure 
of the sky background.

\begin{figure}[ht]
\epsscale{1.0} 
\plotone{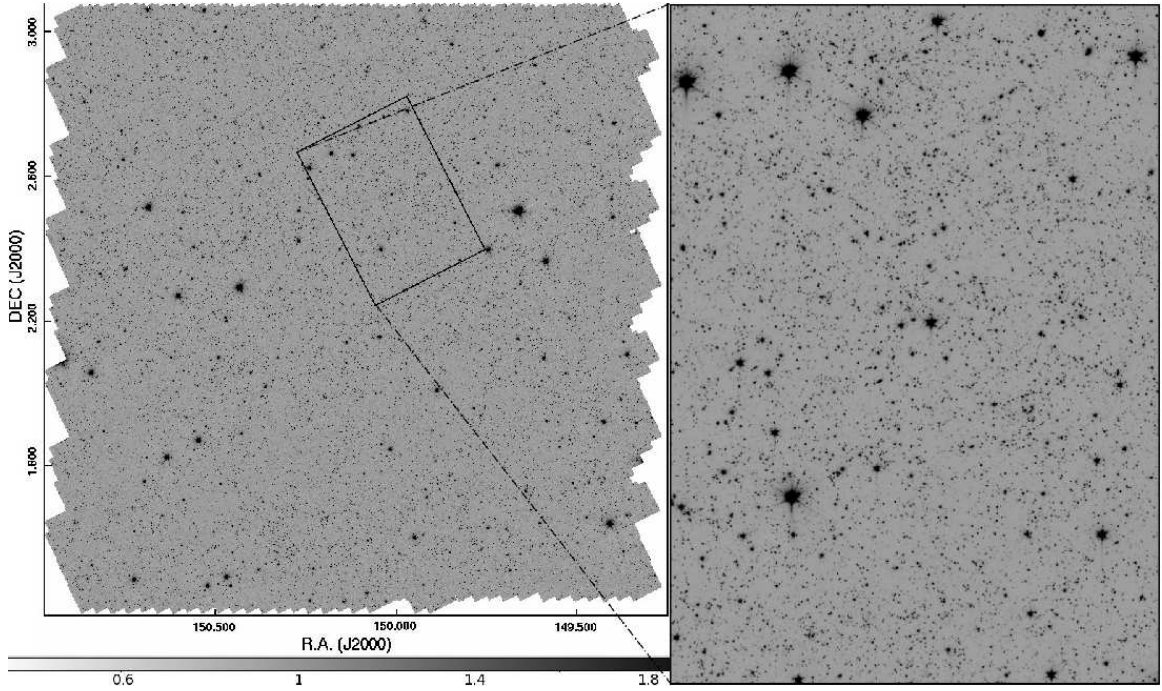}
\caption{{\it Left panel:}\  Mosaic of IRAC 3.6{\ts}$\mu$m Cycle{\ts}2 data for the whole S-COSMOS field. \ \ {\it Right panel:}\ The zoomed area 
corresponding to the location of the MIPS-deep ``Test area" is shown for comparison with the right panel in Figure \ref{image_mips24}.  The intensity scale bar is in units of $\mu$Jy.}
\label{image_irac3.6}
\end{figure}

Once the frame-level images are prepared, they are reprojected to a
common tangent projection and coadded using the SSC MOPEX 
software{\ts}\footnote{http://ssc.spitzer.caltech.edu/postbcd/}
\citep{mak05} to create the final mosaiced image.
An example of the quality of the data can be seen in Figure \ref{image_irac3.6} where the whole 
mosaic is shown (left panel).   The  zoomed area (right panel) better illustrates the 
large space density of 3.6{\ts}$\mu$m sources, and also shows that there are 
no obvious artifacts in the reduced image.

This reprojection also corrects for known image distortion. Since the
BCD images are aligned to 2MASS, the final mosaics are also on the
2MASS astrometric solution, and generally have an accuracy of
$\sim${\ts}0.2\arcsec.  Typical depth of coverage is between $12\times$ and $22\times$
100-second images per point on the sky. This is sufficient redundancy to reject cosmic rays with
very high confidence. In general, asteroids are also rejected (since
they move between observation epochs), although some low-sigma
residuals from slow moving objects remain. These are easily identified by their color.

\subsection{MIPS}

\begin{figure}[ht]
\epsscale{1.0} 
\plotone{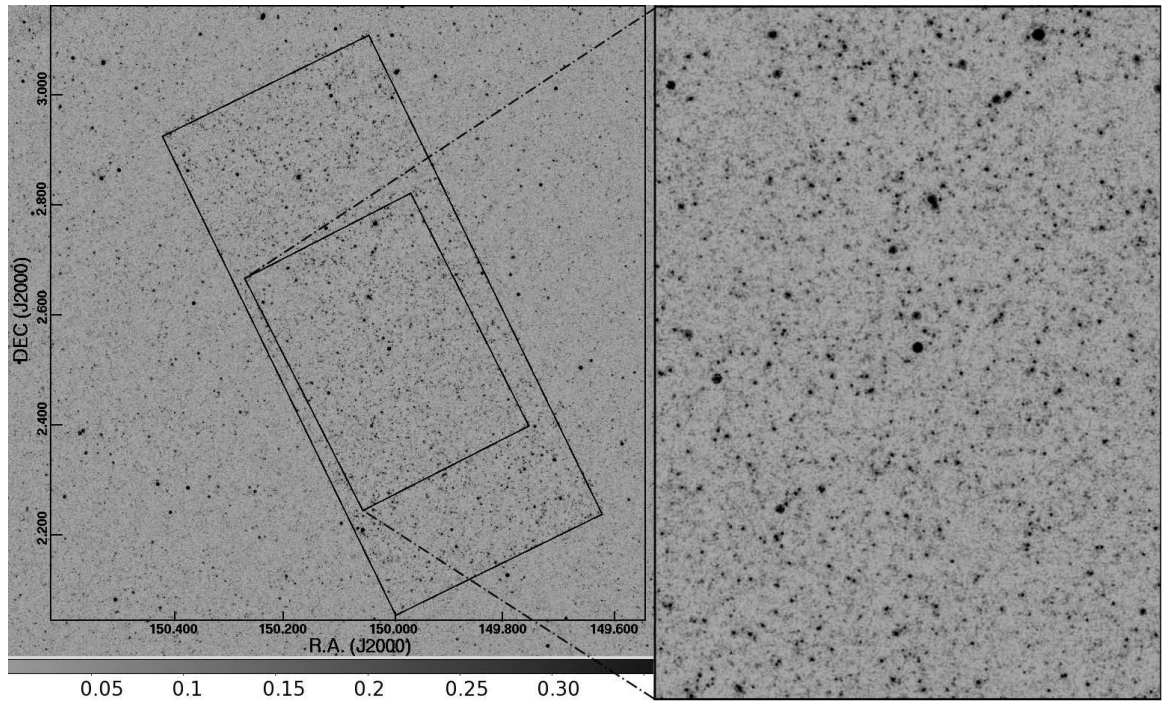}
\caption{\ \ MIPS 24{\ts}$\mu$m sources in the MIPS-deep ``Test area" and larger shallow regions of the S-COSMOS field.  
The large rectangle insert in the left panel represents the full Cycle 2 deep coverage area 
at 24{\ts}$\mu$m,  The smaller rectangle and zoom in the right panel represents the Cycle{\ts}2 area 
with deep coverage in all three MIPS bands. The intensity scale bar is in units of mJy.} 
\label{image_mips24}
\end{figure}

The MIPS 24{\ts}$\mu$m data have been reduced using a combination of
our own developed IDL routines and MOPEX \citep{mak05}. Our starting
point is the BCD products generated by the SSC pipeline version S13.2,
where the data is flux calibrated.  Great attention has been given to
the subtraction of the zodiacal background, the dominant source of
diffuse emission at 24{\ts}$\mu$m at the ecliptic latitude of
S-COSMOS. Our goal was to check for the presence of weak structured
cirrus emission in the COSMOS field: therefore, we avoided using any
background subtraction method that could lower or cancel the signal
from the interstellar medium (ISM).  Given the large size of the
field, the zodiacal background to the detectors varies across the
field, from 36.8{\ts}MJy/sr to 37.8{\ts}MJy/sr, primarily as a
function of the ecliptic latitude, but also as a function of the
ecliptic longitude and of the MIPS scan mirror position. These values
are slightly higher than the ones predicted by SPOT, as seen in
table~\ref{tbl-fieldcomparisons}. This excess of zodiacal background
is due to the presence of dust lanes that are not included in the
model that SPOT uses for its background estimation (W. T. Reach,
private communication).  Following \citet{fad06}, we first subtracted
the variations due to the scan mirror by identifying the sets of
images taken at the same position and then determining the amount of
variation due to the mirror position.  We then subracted the ecliptic
background.  The variations in this background are very smooth, and
therefore do not induce any kind of structure. It was possible to fit
and subtract out this background using a low order polynomial that
would preserve any type of finely structured cirrus-like emission from
the interstellar medium (ISM).  The resulting image is shown in Figure
\ref{image_mips24}.   No cirrus-like structure can be seen in the main
or deep map. We will now quantify this statement, by analysing the
power spectrum of our maps. These are computed following the
prescriptions of \citet{miv02}.

\begin{figure}[ht]
\epsscale{1.0} 
\plotone{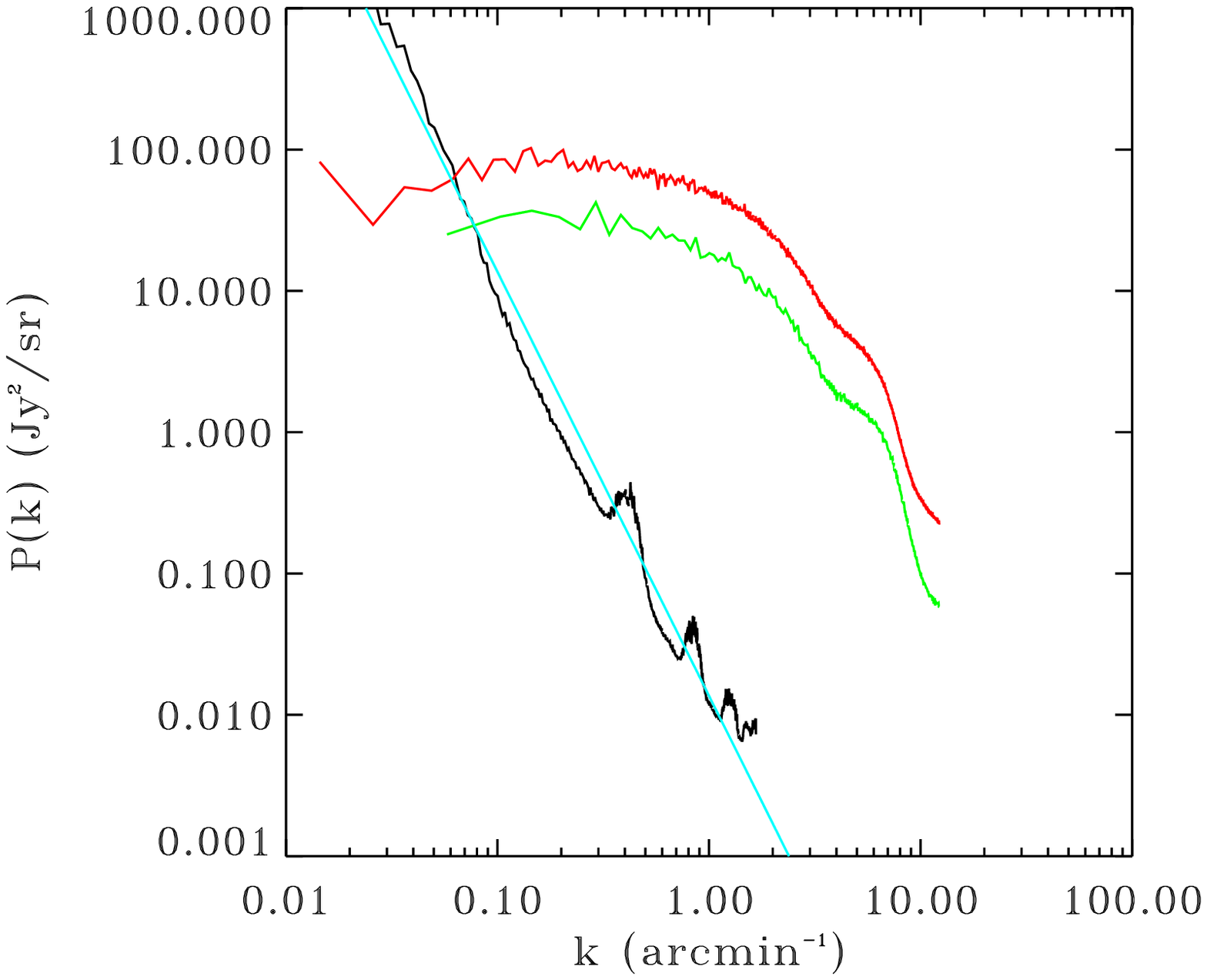}
\caption{\ \ Power spectrum at 24{\ts}$\mu$m in the S-COSMOS area.  Red: power spectrum computed on the MIPS-shallow 24{\ts}$\mu$m image.  Green: power spectrum computed in the MIPS-deep ``Test area".   Black: power spectrum computed on the IRAS 100{\ts}$\mu$m \citet{sch98} image (see Figure~\ref{cosmos_field}), and scaled to 24~$\mu$m using equation \ref{eq:gautier}. Cyan: expected power spectrum from equation \ref{eq:helou}, assuming a power law with exponent $-3$. Both estimations of the cirrus contribution to the power spectrum are in good agreement, and orders of magnitude below the measured power spectrum in the maps, except at the largest scales.}
\label{pk}
\end{figure}

\citet{gau92} have shown that the power spectrum of the cirrus emission of the ISM at 100{\ts}$\mu$m can be approximated by the equation:
\begin{equation}
P(k) \propto B^{2}_{100 \mu m} \left( \frac {k}{k_{0}} \right)^{-3} 
\label{eq:gautier}
\end{equation}
where $k$ and $k_{0}$ are spatial frequencies, $B_{100 \mu{\rm m}}$ is the ISM surface brightness at 100{\ts}$\mu$m  in units of Jy/sr,  and $P(k)$ is the power spectrum in units of Jy$^2$/sr. Note that while the original \citet{gau92} formula had a dependency of $B^3_{100 \mu{\rm m}}$, recent results show that $B^2_{100 \mu{\rm m}}$ is a better representation below a surface brightnesses of 10{\ts}MJy/sr \citep{ing04},  From this behavior, \citet{hel90} have shown that the noise fluctuations due to infrared cirrus in a resolution element are:
\begin{equation}
\frac{N}{1{\ts}{\rm mJy}} = 0.3 \left(\frac{\lambda}{100{\ts}\mu{\rm m}}\right)^{2.5}  \left(\frac{D_{\rm t}}{1{\ts}m}\right)^{-2.5}   \left(\frac{\left<B_{\lambda}\right>}{1{\ts}{\rm MJy/sr}}\right)^{1.5} = \sqrt{P(k) \Omega}
\label{eq:helou} 
\end{equation}
where $\lambda$ is the wavelength of observations, $D_{t}$ is the
telescope diameter, $\left<B_{\lambda}\right>$ is the average surface
brightness of interstellar dust emission at the wavelength of
observations and $\Omega$ the measurement aperture solid angle. The
predicting power of this formula has been verified to be within a
factor of 2 by \citet{kis01} using ISOPHOT observations. In the
S-COSMOS field, we have $\left<B_{100 \mu m}\right> = 0.9${\ts}MJy/sr from
the \citet{sch98} analysis of the IRAS maps. Using typical cirrus
colors from \citet{ing04}, we infer that the contribution of
cirrus to the confusion is expected to be of $N=0.1${\ts}$\mu$Jy at 24{\ts}$\mu$m, 
well below the detection limit of the deep test field. This is
illustrated in Figure \ref{pk} where the power spectrum of the
main and deep test area are plotted respectively in red and
green. Both spectra are dominated above 1{\ts}arcmin$^{-1}$ by the
convolution with the Spitzer-MIPS point spread function that shows an exponential decreases
in the power at small scales.  Note also that the average
ISM emission between the main and deep component is slightly
different, hence resulting in a difference of overall power spectrum
normalization between the two areas. The black curve in
Figure \ref{pk} is the power spectrum measured on the
\citet{sch98} IRAS 100{\ts}$\mu$m, scaled to 24{\ts}$\mu$m using the square
of the color ratio, $S(100\mu{\rm m}) / S(24\mu{\rm m})$,  from \citet{ing04}, as shown
in equation \ref{eq:gautier}. At 100{\ts}$\mu$m, the IRAS map, very
similar to the one displayed in Figure \ref{cosmos_field}, is
dominated by the cirrus structure. Indeed, its power spectrum is very
close to a power law with index $-3$. By transferring it at 24{\ts}$\mu$m,
we obtain an approximation of the power spectrum of the cirrus
component in our map. Another approximation is given by
equation \ref{eq:helou} that gives the rms fluctuations for $k =
10.3$~arcmin$^{-1}$ in the case of Spitzer-MIPS{\ts}24. Assuming a 
power law with index $-3$, we obtain the cyan curve in Figure~\ref{pk}. Note
that we did not convolve the black and cyan curves with the 
Spitzer-MIPS PSF. Doing this would introduce a further exponential
decrease toward high k values as seen in the ones measured on the
maps. Both approximation are in extremely good agreement, and we can
readily see that the cirrus emission is negligeable in the field
expect at the largest scales. Note that the polynomial fitting we
applied to subtract the zodiacal background has suppressed the largest
scales of the map, that are possibly dominated by the cirrus spectrum
below $k=0.07${\ts}arcmin$^{-1}$. The main conclusion from this power
spectrum analysis is that the cirrus emission is negligible at all
scales expect the largest map in the field: its contribution is orders
of magnitude below that of the sources in the field.

Uncertainty maps for both the Main and Deep observations were produced
along the field mosaic. We checked the uncertainty values produced by
the S13.2 pipeline at the BCD level by producing difference images of
overlapping scan legs. These images were used to detect moving objects
(see section \ref{sec:asteroids}). We checked the distribution of pixels
in the difference images and found that the BCD product uncertainties
were over-estimated by 44\%. We corrected all BCD uncertainty products
for this factor before producing the mosaic. Final sensitivities were
derived from the uncertainty maps, assuming a point source and using
the PRF produced by the SSC for MIPS 24{\ts}$\mu$m.

The MIPS-70 and MIPS-160 data were reduced using the
online SSC BCD products following the pipeline algorithms presented in
\citet{gor05} and the filtering techniques of \citet{fra06a}.  We have
adopted the S13 calibration factors (based on stellar SEDs) of
702{\ts}MJy/sr and 44.7{\ts}MJy/sr per MIPS-70 and MIPS-160 units,
respectively, and have applied the appropriate color corrections of
1.09 and 1.04 at 70{\ts}$\mu$m and 160{\ts}$\mu$m respectively to
correct for galaxy SEDs.  With future optimized processing
\citep[e.g.,][]{fra06b}, we hope to improve upon the current S-COSMOS
70{\ts}$\mu$m and 160{\ts}$\mu$m sensitivities
(Table~\ref{tbl-sensitivities}).

\section{Preliminary analysis:  S-COSMOS (Cycle 2) }
\label{prelimanalysis}

\subsection{Backgrounds and Sensitivities}
From our recently completed Cycle-2 observations, we validated the
quality of the field for future deep infrared observations.  As shown
in Table \ref{tbl-fieldcomparisons}, we found total background
levels of 37{\ts}MJy/sr, 10.5{\ts}MJy/sr, and 7.5{\ts}MJy/sr in the
MIPS 24{\ts}$\mu$m, 70{\ts}$\mu$m, and 160{\ts}$\mu$m bands, respectively.
The background is dominated by the zodiacal background at
24{\ts}$\mu$m and 70{\ts}$\mu$m and is remarkably smooth over the
entire COSMOS field.  In addition, no strong cirrus structures were
observed within the field, as demonstrated by our power spectrum
analysis of the 24{\ts}$\mu$m maps.  The zodical background level is $\sim${\ts}$1.8-2\times$
higher than the low background fields on the sky such as the Lockman
Hole (LH), CDF-S, HDF-N, and the Groth Strip, which all have similar
backgrounds (see \citet{lon03} and Table \ref{tbl-fieldcomparisons}).

For an equal amount of integration time, the S-COSMOS MIPS 
sensitivities at 24{\ts}$\mu$m and 70{\ts}$\mu$m are within 35\% of
the lowest background fields, confirming that the MIPS sensitivities
scale as expected with the square root of the background.  At
160{\ts}$\mu$m the COSMOS field has a background level, which is only
15\% higher than the low background fields (only 7\% lower sensitivity
for the same integration time).  Our Cycle 2 MIPS observations of the
full-field (shallow) and the deep ``Test area" (see Table
\ref{tbl-sensitivities}) show that we are able to integrate down as
$\sigma \propto t^{-0.5}$ .  This demonstrates our ability to achieve
our expected sensitivities in the MIPS-deep observations of the full
field to be carried out in Cycle 3.

A complete accounting of the measured sensitivities for our Cycle 2 IRAC and MIPS 
observations of the COSMOS field is given in Table \ref{tbl-sensitivities}. 

\subsection{Preliminary number counts}


\subsubsection{IRAC number counts}

A photometric catalogue has been obtained using the IRAC $3.6\mu$m
mosaic.  The source extraction has been performed using the SExtractor
software \citep{ber96}.  Poor image quality areas have been masked
(e.g. field boundary, a low coverage area on the south-west
corner\footnote{IRAC Coverage of the south-west corner of the COSMOS field 
was partially affected by the loss of 9 AORs due to satellite transmission difficulties
during our Cycle{\ts}2 observations; these AORs are 
scheduled to be recovered in 2006, Dec.}, areas around saturated stars). The final
``effective" area (i.e. after removing the masked regions) is
2.3{\ts}\sq\deg.

Preliminary IRAC galaxy number counts are shown in
Figure \ref{numbercounts_irac}. The number counts shown are the number
of discrete objects detected by SExtractor in the mid-infrared per
flux per area.  Due to the comparably large IRAC beam, comparison with
extremely deep Subaru optical images indicates a significant
difficulty in cross-identification between the optical images and the
infrared images, which affects between 20\% and 30\% of the IRAC
sources. However, what is shown here are strictly infrared number
counts.  Overall, the counts show broad agreement with those taken
from elsewhere in the literature, demonstrating that the preliminary
catalog is substantially free of significant systematics relative to
pre-existing Spitzer surveys.  The S-COSMOS survey is extremely deep,
and photometrically reaches below the classical confusion level of 
40{\ts}beams/source at 3.6{\ts}$\mu$m and 4.5{\ts}$\mu$m. As a result, confusion is significant
at faint levels, and this ultimately limits the completeness of the
counts at $< 1{\ts}\mu$Jy. Below this level the number counts are
systematically underestimated. Because even at the total integration
time of S-COSMOS source confusion is significant, deeper surveys
-- e.g.  IRAC North Ecliptic Pole Dark Field \citep{sur04}, XFLS
ELAIS-N1 \citep{fad04,rod06}, and QSO HS 1700+6416 \citep{faz04,bar04} --  
do not extend to proportionally fainter levels.  Thus, while S-COSMOS is $\sim${\ts}$50\times$
shallower in integration time than those deeper surveys, it reaches
similar flux levels while covering an area between 40 and 200 times
larger.  Thus S-COSMOS is as effective as any existing Spitzer survey
for studying faint galaxy populations, but at the same time minimizes
cosmic variance on one degree size scales when studying those
populations due to it's larger area.  Surveys such as SWIRE will provide more
information on cosmic variance for brighter sources due both to the 
larger total area coverage (25$\times$) as well as sampling multiple 
(6$\times$) fields. However, the SWIRE completeness limit is more than
10$\times$ brighter than that of S-COSMOS, and therefore cannot address the
faintest decade in flux.

\subsubsection{MIPS number counts}

A preliminary catalog has been extracted from the Full-field Shallow
and Deep ``Test area" images.  Source detection has been performed
using the IDL version of the DAOPHOT \citep{ste87} on the S/N images.
This method has been found to be the most efficient in separating the
high ($\sim 25\%$) fraction of blended sources in the deep area where
the depth reaches below 70{\ts}$\mu$Jy.  Source fluxes were measured
on the images using aperture photometry, masking out neighbouring
sources that could contribute significantly. They were then scaled to
total fluxes, using the MIPS{\ts}24{\ts}$\mu$m Point Response Function
(PRF) provided by the SSC, and taking the masked pixels into
account. This method allows us to deal with the blending in a
preliminary way.

Stars were identified using both the \citet{rob06} catalog derived from
ACS morphology, and the 2MASS PSC catalog \citep{skr06} for the bright
end. Additionally, we checked against our XMM point source catalog in
order not to include any AGN in the star-like objects. We find that
stars account for roughly 5\% of the MIPS 24{\ts}$\mu$m detections at 1
mJy. This fraction rises to 50\% at the bright end of the counts.

The number counts for both the S-COSMOS full-field MIPS-shallow and
MIPS-deep ``Test area" images are displayed in Figure
\ref{numbercounts_mips}, together with the results from the GTO
surveys compiled by \citep{pap04}. The counts from the deep area agree
well at their bright end with the ones from the full-field at their
faint end. They also agree within the Poisson error bars with the GTO
counts. The effect of stellar contamination is clearly visible in the
counts on the MIPS shallow area where both the counts with and without
the inclusion of stars is shown.

 \subsection{Asteroids\label{sec:asteroids}}

 In the context of number counts, asteroids are considered a
 contaminant and are masked in the final image.  On the other hand,
 {\it Spitzer} is ideally suited to study the thermal emission,
 especially from Main Belt Asteroids (MBAs), whose peak flux is
 typically measured at $15-20${\ts}$\mu$m.  Understanding the
 current Size Frequency Distribution (SFD) of the MBA population
 provides important insight into the evolution of the solar system,
 including accretion and collisional history, as well as impact
 hazards to the Earth.  Although \citet{ted05} have suggested
 that the MBA population is composed of a significant fraction of
 smaller (sub-kilometer diameter) objects, these bodies have only been observable since
 the launch of Spitzer.  Their Figure 6, based on ground-based observations of $\sim 1.9 \times
10^6$ known asteroids and extrapolation of the SFDÕs of several
asteroid families, suggests that the number of objects increases
logarithmically with decreasing diameter.  The actual number of
smaller objects has been difficult to empirically determine, however,
as MBA's emit most of their radiation in the mid-IR, leading to
observational bias towards larger, more reflective asteroids.

Figure \ref{asteroid_lameia_sed} shows the observed optical and IR SED of 
asteroid MBA{\ts}248, viewed at heliocentric and Spitzer-centric distances 
of 2.63{\ts}AU and1.62{\ts}AU, respectively.  The optical and IR emission is from 
reflected sunlight and thermal emission, respectively. The MBA{\ts}248 SED 
can be considered typical of asteroid SEDs at these wavelenghts.   
As predicted by the widely used so-called
``Standard Thermal Model" \citep{leb89}, these solar system bodies
are expected to emit substantially more radiation in the mid-IR than
at optical wavelengths, thus making ground-based detection feasible for
only a subset of the MBA population.

S-COSMOS, carried out near the ecliptic plane at $\beta \sim -9^\deg$,
has detected more than 120 asteroids per \sq\deg\ at
24{\ts}$\mu$m, down to a sensitivity limit of 0.30{\ts}mJy ($5\sigma$), in the 
Cycle{\ts}2 MIPS-deep ``Test area".  For comparison, the only Spitzer
asteroid survey carried out to date, the First Look Survey Ecliptic
Plane Component (FLS-EPC), has looked at number counts in the ecliptic
plane (i.e. $\beta \sim 0^\deg$),  from 8{\ts}$\mu$m IRAC observations down to a sensitivity limit 
of 0.08{\ts}mJy ($5\sigma$), and these data indicate that $125 \pm 33$ 
asteroids per \sq\deg\ are present.

Because the SFD for the MBA population is expected to change with distance from the ecliptic
plane \citep{bro03}, and the sensitivity limits of the two programs
are different, a direct comparison between S-COSMOS and FLS-EPC
asteroid number counts cannot be made. The FLS-EPC has shown that
the asteroid mid-IR color, $S(24\mu{\rm m}) / S(8\mu{\rm m})$, appears to be constant over a
large range of diameters and albedos \citep{bha04,mea04}, and this result can be used to reconcile the
two datasets.  As the FLS-EPC suggests that $S(24\mu{\rm m}) / S(8\mu{\rm m})
\sim 10$ down to a sensitivity limit of $\sim$0.08{\ts}mJy ($5\sigma$), a relevant comparison
to S-COSMOS MIPS-deep ``test area" asteroids should only be made for
objects with $S(24\mu{\rm m}) \geq 0.8${\ts}mJy, with latitudinal dropoff
modeled by a power law \citep{cel91}.   As part of our future
work, we plan to carry out this analysis, as well as detailed
photometric analysis and thermal modeling, to refine overall number
counts as discussed above, as well as to refine color values and to
determine orbits.  It is anticipated that the Cycle{\ts}3 MIPS-deep
observations of the full COSMOS field will detect
$\sim${\ts}12{\ts}$\times$ the number of asteroids compared to what
were found in the Cycle{\ts}2 MIPS-deep ``Test area", thus reducing
the uncertainty in the preliminary number counts per \sq\deg\ provided here.

\section{Data Quality and Future Science Goals}
\label{future}

The preliminary results from our S-COSMOS Cycle{\ts}2 IRAC and MIPS
observations have confirmed expectations that the COSMOS field is one
of the best equatorial fields of comparable size available for
carrying out deep infrared surveys.  Our preliminary analysis has
demonstrated that (1) the full COSMOS field is devoid of cirrus
contamination, (2) we are able to achieve our predicted sensitivities
with both IRAC and MIPS cameras within the allotted integration times,
(3) measured background levels were very close to those predicted,
demonstrating our ability to minimize the zodical background levels
through judicious use of the visibility windows, and (4) our observing
methods allow for efficient removal of asteroid ``contaminants", and
in fact have resulted in the discovery of many new Main Belt
Asteroids.

The IRAC and MIPS sensitivities achieved in Cycle{\ts}2, (and expected
for Cycle{\ts}3 MIPS), will allow us to carry out the two main science
goals of S-COSMOS: a study of the stellar-mass assembly of galaxies
and a full accounting of the luminosity from dust-embedded sources
such as merging starburst galaxies and AGN, primarily in the redshift
range $z \sim 0.5-3$ where COSMOS is designed to study galaxy
evolution as a function of large-scale structure environment and
redshift
\citep{sco06}.   

\begin{figure}[ht]
\epsscale{0.8} 
\plotone{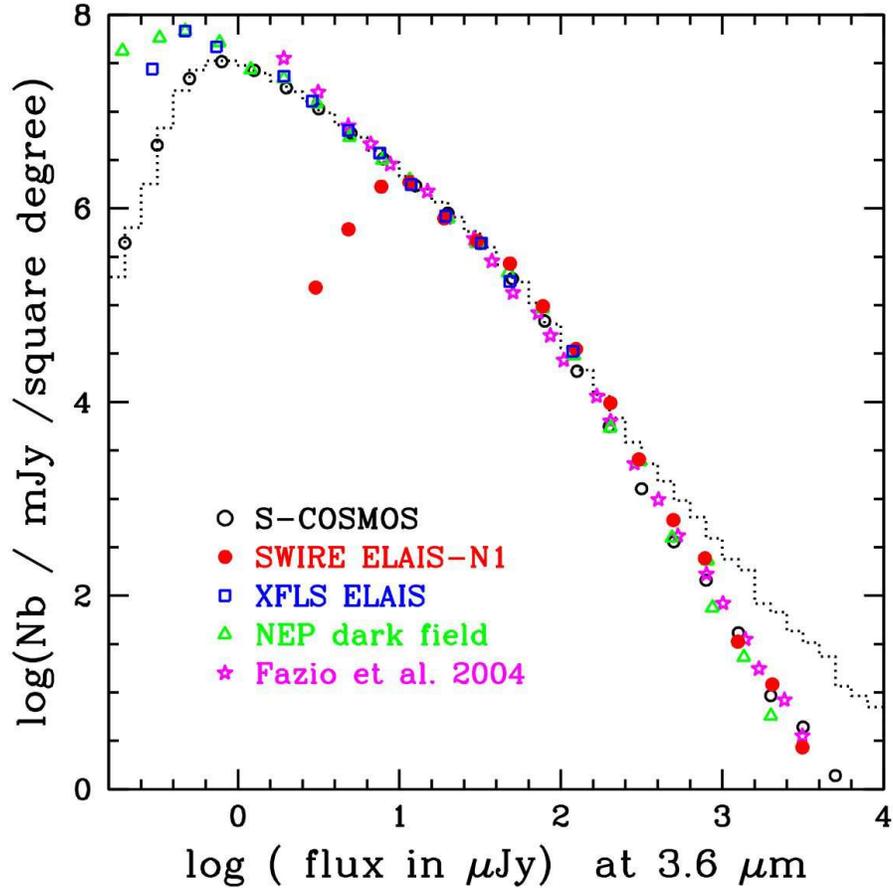}
\caption{\ \ Differential number counts at 3.6{\ts}$\mu$m (IRAC-1) for the full-field Cycle 2 S-COSMOS observations.
 The stars are removed out to $i'_{AB}<23$ by using a morphological
 criterion, FWHM/surface brightness, as measured in the HST/ACS image
\citep{cap06}.  Given that AGN appearing as quasi stellar objects could also be 
removed by this criterion, we checked and found that X-ray selected AGN
represent only 2\% of the removed population; thus adding them back to
the galaxy sample has only a small impact on the galaxy number counts.
Black open circles and dotted line represent the S-COSMOS counts
without and with the stars, respectively (see text).  The SWIRE
ELAIS-N1, XFLS ELAIS, and NEP dark field data are from \citet{sur07}.
None of the number counts have been corrected for
incompleteness.}
\label{numbercounts_irac}
\end{figure}

\begin{figure}[ht]
\epsscale{0.8} 
\plotone{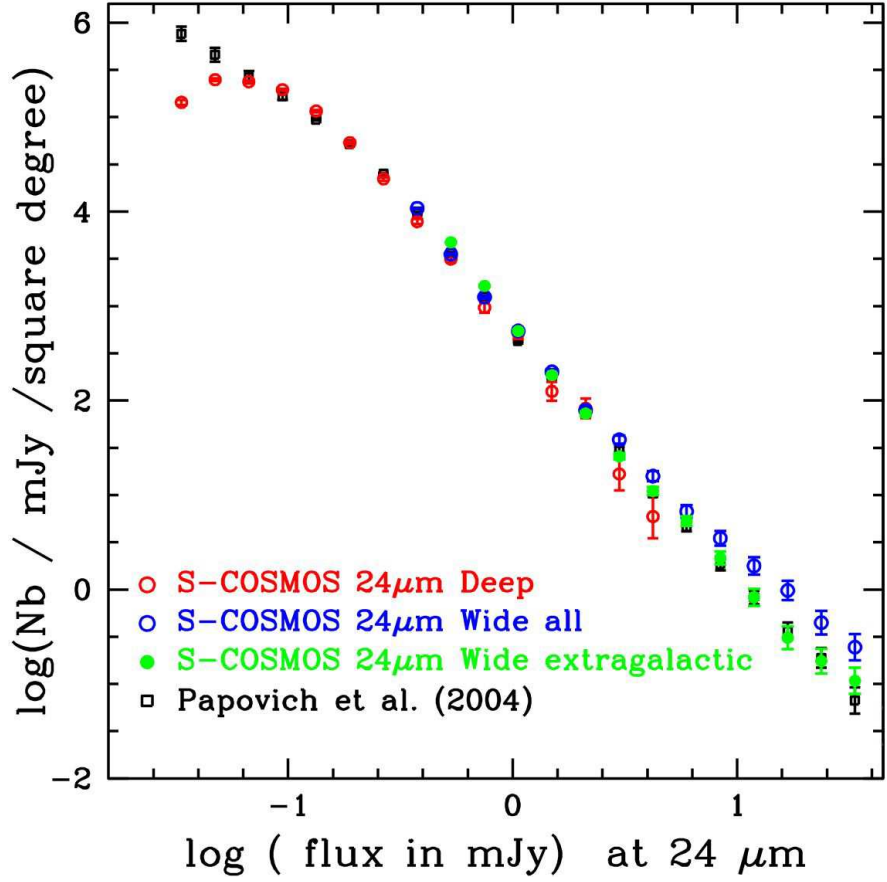}
\caption{\ \ Preliminary differential counts at 24{\ts}$\mu$m from the MIPS-deep ``Test area" (red) and 
MIPS-shallow ``Wide all" (cyan) Cycle{\ts}2 S-Cosmos observations, compared to those compiled by \citet{pap04}. 
The error bars for the S-COSMOS counts are the $1 \sigma$ poissonian fluctuations. The counts in the MIPS-deep 
area are compatible with the Papovich counts down to our $5 \sigma$ completeness limit of $67 \mu$Jy. 
In the MIPS-shallow ``Wide all" observations, the excess of sources detected at large fluxes is due to stars in the field.  
The MIPS-shallow ``Wide extragalactic" (green) counts were obtained after removal of the stars (see text).} 
\label{numbercounts_mips}
\end{figure}

\begin{figure}[ht]
\epsscale{1.0} 
\plotone{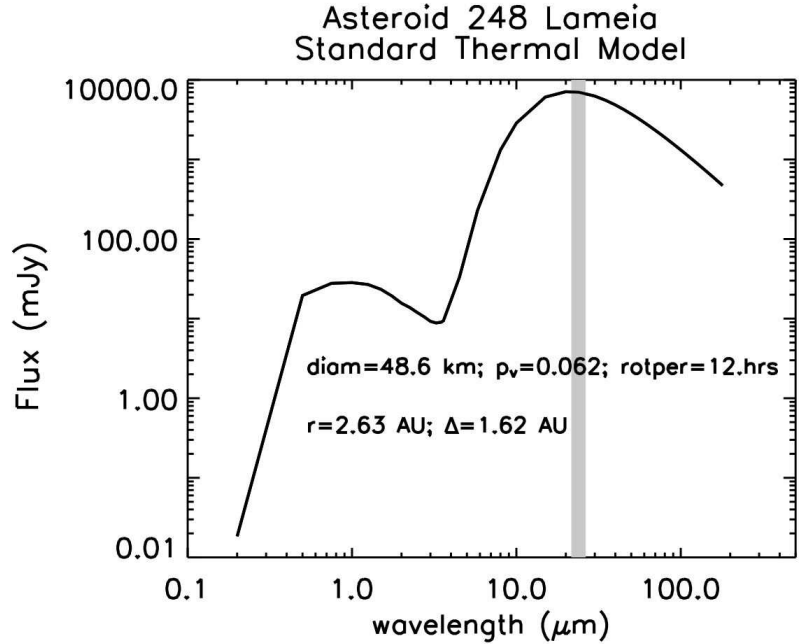}
\vskip -0.4truein
\caption{Model spectral energy distribution for main belt asteroid 248{\ts}Lameia.  The Standard Thermal Model, assuming geometric albedo pv = 0.062 is used.  As this object is highly non-reflective, the mid-IR emission is two orders of magnitude greater than at optical wavelengths.  The grey bar represents the MIPS 24{\ts}$\mu$m  band.  This object's thermal properties are similar to those of other MBA's, making them ideally suited for detection at the S-COSMOS wavelengths of 8{\ts}$\mu$m and 24{\ts}$\mu$m.}
\label{asteroid_lameia_sed}
\end{figure}

An estimate of the numbers of galaxies that will be detected by
S-COSMOS can be made using the SEDs shown in Figure \ref{galaxy_seds}, 
along with the local LF for infrared-selected galaxies \citep[e.g.,][]{sanea03} 
and a model for evolution of the LF with redshift \citep[e.g.,][]{kim98,san03}.  
Figure \ref{galaxy_seds} shows four
redshifted galaxy SEDs together with the measured sensitivities for
the S-COSMOS and HST-ACS surveys.  Included are a massive
bulge-dominated system (``M87-like"), an actively star-forming disk
(``M51-like"), and both ``cool" and ``warm" ultra-luminous IR/merging
systems (e.g. Arp 220, Mrk 231) with varying mixtures of starburst and
AGN activity.  Our integration time of $\sim${\ts}1200{\ts}sec per
pixel in each of the four IRAC bands is sufficient to easily detect
$\sim${\ts}$L^*$ disks and spheroids out to $z \sim${\ts}3.  The total
number of objects detected by IRAC over the full COSMOS is  
expected to be $\gtrsim 10^6$ as determined from the differential number
counts at 3.6{\ts}$\mu$m (see Figure \ref{numbercounts_irac}).  The
MIPS-deep integration times of $\sim${\ts}3000{\ts}sec at
24{\ts}$\mu$m will allow us to detect luminous infrared galaxies
(LIRGs: $L_{\rm ir} > 10^{11} \Lsol$) out to $z \sim${\ts}2, and
ultraluminous infrared galaxies (ULIRGs: $L_{\rm ir} > 10^{12} \Lsol$)
out to $z \sim${\ts}3.  To compute the total number of objects
expected to be detected by MIPS, we have assumed strong evolution 
of the form $(1+z)^{5.5}$ for objects with $L_{\rm ir} \gtrsim 3\times10^{11} L_\odot$, consistent 
with the strong evolution either measured or inferred from previous studies 
\citep[e.g.,][]{san03,oya05,lef05} which show that the co-moving space 
density of the LIRG/ULIRG population appears to increase by 2--3 orders of magnitude 
over the redshift range 0--2.  At 24{\ts}$\mu$m S-COSMOS should detect $\gtrsim${\ts}$10^5$ LIRGs out to
$z \sim 2$ and $\gtrsim${\ts}$3\times10^3$ ULIRGs out to $z \sim 3-4$.
At 70{\ts}$\mu$m and 160{\ts}$\mu$m S-COSMOS should detect
$\sim${\ts}$10^3$ ULIRGs out to $z \sim 1.5-2$.

S-COSMOS is optimal for selecting unbiased samples of AGN --almost
independently of their level of obscuration.  At mid-infrared
wavelengths the obscuring dust that hides AGN from ultraviolet,
optical, and soft X-ray surveys should be a strong and largely isotropic
emitter.  As shown by \citet{lac04} and \citet{ste05} and references therein, selection based
on mid--IR colors ( specifically, IRAC and MIPS colors from 3.6
$\mu$m to 24$\mu$m) can not only help disentangle 
stars, galaxies and AGN, but can also help separate type{\ts}1 and type{\ts}2 AGN using the
same criteria. This is crucial if one wants to quantify the fraction
of type2/type1 in the universe.  In fact, obscuration models that aim
to resolve the hard ($>${\ts}8 keV) X-ray background (XRB) debate whether
the ratio remains constant during AGN evolution \citep[e.g.,][]{tre05}, or if it 
varies as a function of redshift and luminosity \citep[e.g.,][]{gil06}.  
Thus, the COSMOS field, with its size, deep mid-IR coverage, and extensive
UV/X-ray/Radio/Optical data set offers not only the possibility to
significantly improve the constraint on the surface density of
obscured AGN, but also to improve our understanding of their spectral
energy distributions (SEDs), and hence the ratio of type2/type1.

The S-COSMOS data will be combined with the full multi-wavelength
COSMOS data set to compute fundamental properties of galaxies.
IRAC+MIPS data will provide a much more complete accounting of the
bolometric luminosity, particularly for luminous starbursts and dusty
AGN.  In addition, the combination of IRAC (e.g. $5-8${\ts}$\mu$m) and
MIPS (e.g. 24{\ts}$\mu$m) data will be essential in order to provide a
useful discriminant between starburst and AGN activity
\citep[e.g.,][]{rig04,pro04,ste05}.  The S-COSMOS IRAC and MIPS colors
can also be used as an AGN versus star formation discriminator in
faint radio sources \citep{sch06}.  Once a galaxy has been identified
as actively forming stars, the combination of radio and Spitzer data,
as well as the relatively accurate photometric, and in some cases
spectroscopic redshifts, will be used to both calibrate and determine
the cosmic evolution of the radio--FIR correlation for star forming
galaxies.  The S-COSMOS observations will also be combined with the
deep radio and (sub)millimeter observations of the COSMOS field to
determine accurate radio through rest-frame far-infrared SEDs of the
sub-millimeter galaxies in the COSMOS field \citep{ber06}.  These
observations will provide the best constraints to date on the physical
conditions in the sub-millimeter galaxies, (e.g. dust temperatures,
dust masses, and bolometric luminosities).

The S-COSMOS data will also be used to select different populations of
new and interesting objects for further study.  For example, 
using our IRAC S-COSMOS data in combination with our optical and
near-infrared COSMOS data, ULIRGs can be effectively selected by the
$BzK$ criterion up to $z \sim 2.5$ \citep{dad05}, whereas a similar
criterion based on the $RJL$ bands (hence exploiting the IRAC
3.5{\ts}$\mu$m data from S-COSMOS) will expand the selection of these
objects up to $z \sim 4$
\citep{dad04}.  Also, for a typical Type{\ts}1 AGN SED, both our IRAC
and MIPS S-COSMOS observations will detect objects with optical
luminosity $1.5 \times 10^{44}${\ts}ergs/s, twenty times fainter than
the mean luminosity of SDSS QSOs at a typical redshift of $z = 1.5$
\citep{ric06}.

It is clear that S-COSMOS will provide a critical part of the overall
COSMOS study of galaxy evolution at $z \sim 0.5-3$, particularly at $z
> 1$ where luminous infrared galaxies become a dominant population of
extragalactic objects.  Interactions and mergers are very likely
responsible for driving the strong evolution seen in the LIRG/ULIRG
population over the redshift range $z \sim 0 - 2$
\citep[e.g.,][]{san03,lef05,bab06}, and therefore galaxy environment
will almost certainly play a prominent role in triggering these
systems \citep{kau03, kau04}.  One of the primary goals of S-COSMOS
will be to explore the role of environment in producing the LIRG/ULIRG
population at $z = 0.5 - 2.5$, and using HST imaging, to determine the
galaxy types involved in the process.

The S-COSMOS Cycle{\ts}2 MIPS and IRAC reduced images and catalogs will 
made publicly available in May, 2007, while the Cycle{\ts}3 MIPS reduced images and 
catalogs will  be made public in June, 2008, one year after the final S-COSMOS 
Cycle{\ts}3 MIPS observing campaign.   A more complete description of the 
data release products and timeline can be found on the Spitzer Legacy Program 
web page $<$http://ssc.spitzer.caltech.edu/legacy/all.html$>$

\acknowledgments
 
 This work is based on observations made with the Spitzer Space
 Telescope, which is operated by the Jet Propulsion Laboratory,
 California Institute of Technology under NASA contract 1407.  Support
 for this work was provided by NASA through Contract Number 1278386
 issued by JPL.  It is a pleasure to acknowledge the hospitality
 provided by the Aspen Center for Physics where the majority of this
 paper was written.  Additional information on the S-COSMOS Legacy
 survey is available from the main COSMOS web site at 
 $<$http://www.astro.caltech.edu/cosmos$>$. It is a pleasure to
 acknowledge the excellent services provided by the NASA IPAC/IRSA
 staff (Anastasia Laity, Anastasia Alexov, Bruce Berriman and John
 Good) in providing online archive and server capabilities for the
 COSMOS datasets. We thank Guilaine Lagache for useful comments on
 power spectra measurments.  We would also like to thank the referee for 
 comments and suggestions which helped clarify our presentation.

 
 
 {\it Facilities:} \facility{HST (ACS)}, \facility{Spitzer (IRAC)}, \facility{Spitzer(MIPS)}.
 
 
 


 \clearpage

 
\begin{figure}[h]
\epsscale{1.0}
\plotone{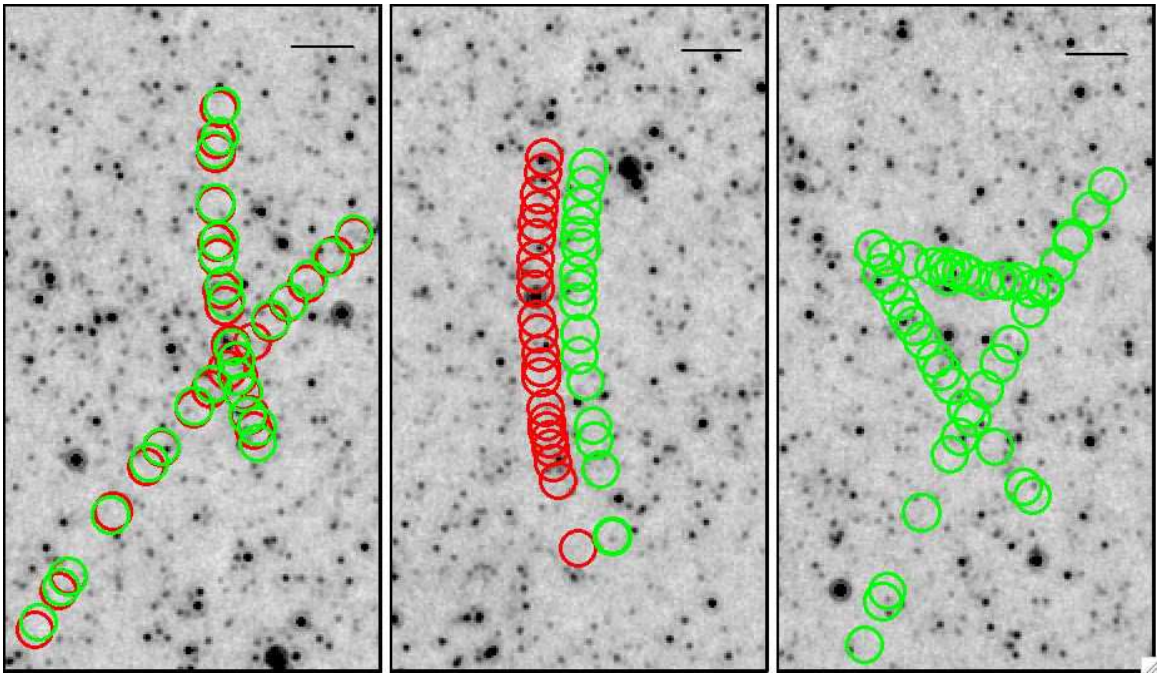}
\caption{Example of asteroids and their trajectories (green circles) on the deep 
MIPS 24{\ts}$\mu$ image.  The figure has the asteroids removed.  Green
circles reflect the original position of the removed asteroids.  We
queried the JPL Horizons On-Line Ephemeris Service using a
Spitzer-centric line of sight (http://ssd.jpl.nasa.gov/x/ispy.html) in
order to quantify the number of asteroids already known.  Red circles
reflect expected locations for the asteroids based on these queries.
Less than 20\% are known to the system (red circles in the first and
middle panel) while 80\% are new detections (right panel).  Among the
asteroids already known, some have a trajectory that is only
approximately known, therefore an offset is measured when compared
with our detections (middle panel).  The scale bar in the upper right
of each panel represents 1\arcmin\ .}
\label{asteroid_tracks}
\end{figure}

\begin{figure}[hb]
\epsscale{1.0} 
\plotone{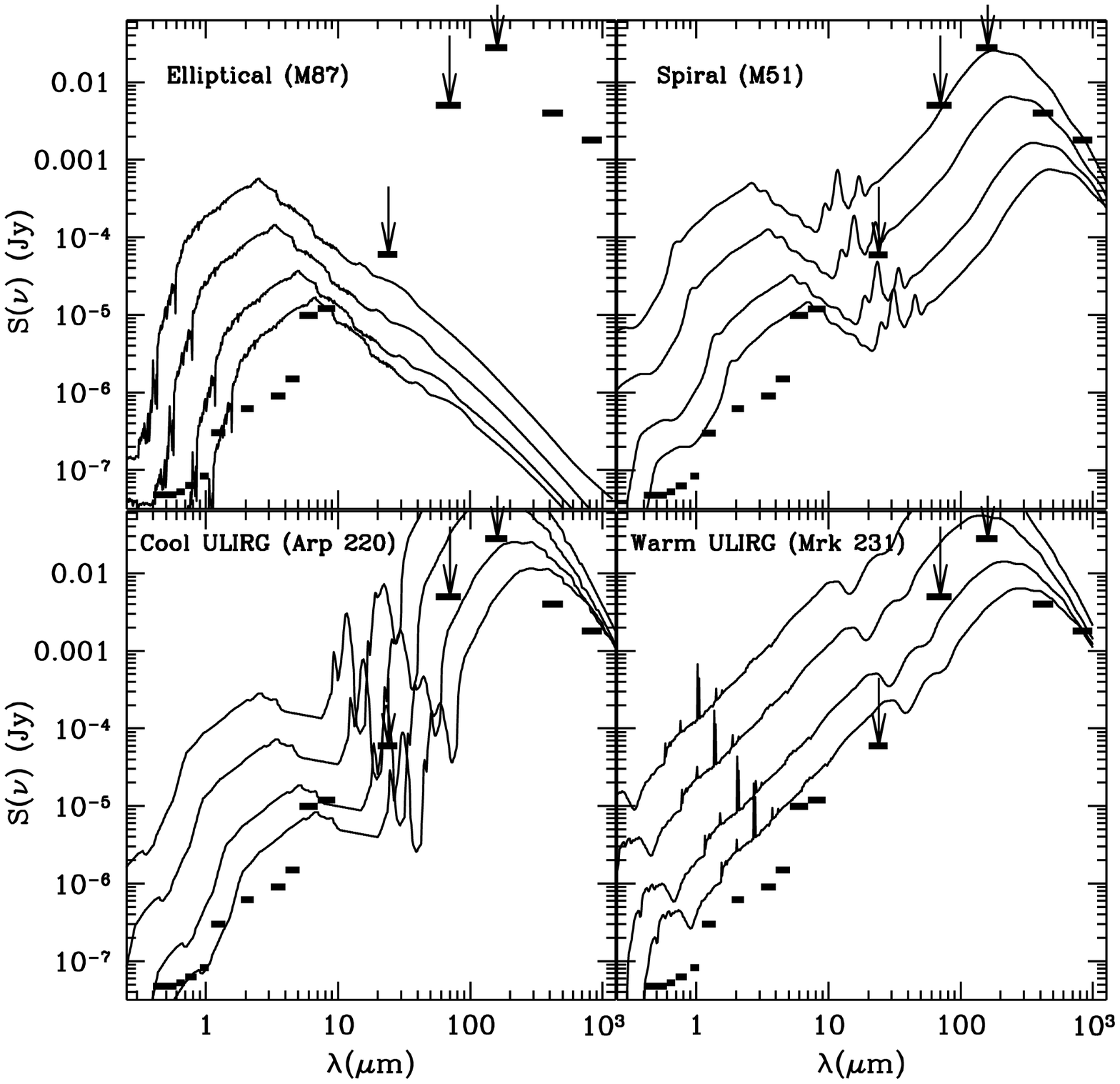}
\caption{SEDs of galaxies \citep{fra01}, as a function of $z$ (= 0.5, 1, 2, 3) (solid curves, top-to-bottom).  
The Elliptical (``M87-like") and Spiral (``M51-like") templates have been scaled to 
$L_{\rm bol} =10^{11} \Lsol$ and $L_{\rm bol} = 3\times 10^{11} \Lsol$ respectively.  
The ``cool" and ``warm" ULIRG templates have been scaled to $L_{\rm bol} = 2 \times 10^{12} \Lsol$.
Superimposed in color are S-COSMOS and COSMOS survey sensitivities ($5\sigma$): \  (red arrows and bars)
S-COSMOS Cycle{\ts}2 MIPS-shallow and Cycle{\ts}2 MIPS-deep surveys, respectively (this paper);  (magenta bars) 
S-COSMOS Cycle{\ts}2 IRAC survey (this paper);  (blue bars) COSMOS ground-based, optical/NIR 
(BVr$^\prime$i$^\prime$z$^\prime$JK) data \citep{cap06,tan06}.   
The purple bars at 450{\ts}$\mu$m and 850{\ts}$\mu$m show the expected sensitivities ($5\sigma$) from our recently 
approved UH+Tri-national Legacy survey of the COSMOS field with JCMT-SCUBA2 \citep{hol03}.} 
\label{galaxy_seds}
\end{figure}

 \clearpage
 
\begin{deluxetable}{c}
 \tablecaption{S-COSMOS (Cycle 2) Observations Summary Table\label{tbl-obssummary}}
 \tablewidth{0pt}
 \tablehead{
 } 
 \startdata
{\bf IRAC-deep Scan Map Summary} \\ 
Three epochs, separated by $4-5${\ts}hrs each (1-15 Jan, 2006) \\ 
Twelve pointings of 100{\ts}sec each\\ 
Total depth = 1200{\ts}sec\\ 
Total time = $166${\ts}hrs\\ 
\\
{\bf MIPS-shallow Scan Map Summary}\\
One epoch (Jan,  2006)\\
Total Depth: 80{\ts}sec (24{\ts}$\mu$m), 40{\ts}sec (70{\ts}$\mu$m), 8{\ts}sec (160{\ts}$\mu$m)\\
Total time = 16.6{\ts}hrs\\
\\
{\bf MIPS-deep ``Test area"  Scan Map Summary}\\
One epoch (Jan,  2006)\\
Total Depth: 3200{\ts}sec (24{\ts}$\mu$m), 1560{\ts}sec (70{\ts}$\mu$m), 320{\ts}sec (160{\ts}$\mu$m)\\
Total time = 41.6{\ts}hrs\\

\enddata
\end{deluxetable}
  
\clearpage

\begin{deluxetable}{lcccc}
\tablecaption{Coordinates for S-COSMOS Observations in  Cycle{\ts}2 and Cycle{\ts}3 \tablenotemark{a}}\label{tbl-fieldscoo}
\tablewidth{0pt} 
\tablehead{
\colhead{Field} &\colhead{Botton Left} &\colhead{Top Left}& \colhead{Top Right}&\colhead{Botton Right}
}
\startdata
IRAC 3.6/5.8&10:03:45 1:23:14&10:03:45 3:04:43 &09:57:10 3:04:43&09:57:11 1:23:14\\
IRAC 4.5/8.0&10:03:36 1:17:38&10:03:36 2:58:53 &09:57:03 2:58:53&09:57:03 1:17:38\\
MIPS-deep          & 10:00:14 2:14:35 &10:01:04 2:40:11&09:59:53 2:49:29&09:59:02 2:23:13\\
MIPS-shallow      &10:02:10 0:48:57&10:06:07 2:45:33&09:58:53 3:39:43&09:54:59 1:42:34\\
 & \\
\hline
 & \\
MIPS 24 (1) &10:02:34 1:00:37&10:04:54 2:49:46&09:58:24 3:18:15&09:55:57 1:43:03\\
MIPS 24 (2) &10:02:04 0:59:13&10:05:20 2:42:46&09:59:08 3:26:25&09:56:06 1:29:17\\
MIPS 70 (1) &10:02:47 1:11:06&10:05:05 2:59:06&09:58:41 3:30:09&09:56:21 1:41:25\\
MIPS 70 (2)&10:01:45 0:48:43&10:04:58 2:31:48&09:58:47 3:15:13&09:55:35 1:31:37\\
MIPS 160 (1)&10:02:26 0:54:47&10:04:40 2:39:16&09:58:13 3:09:09&09:55:59 1:23:13\\
MIPS 160 (2)&10:02:23 1:09:14&10:05:29 2:48:08&09:59:18 3:31:47&09:56:12 1:51:41\\
\enddata
\tablenotetext{a}{Cycle{\ts}2 coordinates are given in rows 1--4.  Cycle{\ts}3 coordinates are given in rows 5--10, 
where (1) and (2) refer to the different spacecraft orientations in Jan, and May, 2006, respectively.  The layout of 
the Cycle{\ts}2 and Cycle{\ts}3 maps can be seen in Figure \ref {coverage_cycle2} and Figure \ref {coverage_cycle3}, respectively}
\end{deluxetable}

\clearpage  

 \begin{deluxetable}{lccccc}
 \tablecaption{Field Comparison of IR Backgrounds, and Sensitivities with Predicted and Measured S-COSMOS Cycle-2 IRAC-deep and MIPS-deep Observations.\label{tbl-fieldcomparisons}}
 \tablewidth{0pt}
 \tablehead{
     & \multicolumn{2}{c}{8.0{\ts}$\mu$m (spot)}  &  \multicolumn{2}{c}{24{\ts}$\mu$m (spot)}   &  \colhead{100{\ts}$\mu$m ({\it IRAS})} \\
 \colhead{Field} &  \colhead{Bkg} &  \colhead{{\it S}(1200{\ts}sec)} & \colhead{Bkg} & \colhead{{\it S}(3200{\ts}sec)} & \colhead{Bkg}\\
     & \colhead{(MJy/sr)} & \colhead{($\mu$Jy)} & \colhead{(MJy/sr)} & \colhead{(mJy)} & \colhead{(MJy/sr)} } 
 \startdata
COSMOS\tablenotemark{a} & 6.7  & 14.6 & 31 & 0.065 & 0.90 \\ 
LH, CDF-S\tablenotemark{a} & 5.2 & 12.7 & 19  & 0.053 & 0.45 \\ 
SWIRE-XMM\tablenotemark{a} & 7.1   & 14.8 & 31  & 0.065 & 1.25\\ 
\hline
COSMOS\tablenotemark{b} & 6.9 & 14.6 & 37 & 0.071 & 0.90 \\
 \enddata 
 \tablenotetext{a}{Total backgrounds predicted by SPOT at 8{\ts}$\mu$m and 24{\ts}$\mu$m are given for the different fields in the first three rows.  Sensitivities (5$\sigma$) were computed assuming 1200{\ts}sec of integration with IRAC and 3200{\ts}sec with MIPS, equivalent to our Cycle{\ts}2 integration times for IRAC and MIPS-deep, respectively.  All {\it IRAS} 100{\ts}$\mu$m values are actual measurements from the {\it IRAS} maps.}
 \tablenotetext{b}{Measured median background levels over the full 2{\ts}\sq\deg\  COSMOS field in our S-COSMOS Cycle-2 data.}
  \end{deluxetable}

\clearpage

    \begin{deluxetable}{lcccc}
 \tablecaption{S-COSMOS (Cycle 2) Integration Times and Measured Sensitivities\label{tbl-sensitivities}}
 \tablewidth{0pt}
 \tablehead{
 \colhead{Camera} & \colhead{Band} & \colhead{Coverage} & \colhead{Int. time\tablenotemark{a}} & \colhead{$5\sigma$} \\
   &  \colhead{$\mu$m} & \colhead{(\% field)} &\colhead{(sec)} & \colhead{(mJy)} } 
 \startdata
IRAC & 3.6   & 100 &1200 & 0.0009 \\ 
IRAC & 4.5   & 100 &1200 & 0.0017 \\ 
IRAC & 5.6   & 100 &1200  & 0.0113\\ 
IRAC & 8.0   & 100 &1200 & 0.0146\\ 
MIPS-shallow & 24  & 100 & 80 & 0.42\\ 
MIPS-shallow & 70  & 100 & 40 & 34.0\\ 
MIPS-shallow & 160  &100 & 8 & 150.0 \\ 
MIPS-deep & 24   & 8 & 3200 & 0.071\\ 
MIPS-deep & 70   & 8 & 1560 & 7.5\\ 
MIPS-deep & 160   & 8 & 320 & 70.0\\ 
 \enddata
\tablenotetext{a}{Median effective integration time per pixel.}
\end{deluxetable}

 \end{document}